\documentclass{amsart}
\usepackage{amssymb}
\usepackage{graphicx}
\usepackage{amscd}

\newtheorem{theorem}{Theorem}
\theoremstyle{plain}

\newtheorem{corollary}{Corollary}

\newtheorem{lemma}{Lemma}

\numberwithin{equation}{section}

\begin{document}
\title{A functional Generalized Hill process and applications}
\subjclass[2010]{Primary 62E20, 62F12, 60F05. Secondary 60B10, 60F17}
\keywords{Extreme values theory; Asymptotic distribution; Functional
Gaussian and nongaussian laws; Uniform entropy numbers; Asymptotic
tightness, Stochastic process of estimators of extremal index; Sowly and
regularly varying functions.}

\begin{abstract}
We are concerned in this paper with the functional asymptotic behaviour of
the sequence of stochastic processes 
\begin{equation}
T_{n}(f)=\sum_{j=1}^{j=k}f(j)\left( \log X_{n-j+1,n}-\log X_{n-j,n}\right) ,
\end{equation}
indexed by some classes $\mathcal{F}$ of functions $f:\mathbb{N} \backslash \{0\} \longmapsto \mathbb{R}_{+}$ and where $k=k(n)$ satisfies 
\begin{equation*}
1\leq k\leq n,k/n\rightarrow 0\text{ as }n\rightarrow \infty .
\end{equation*}
This is a functional generalized Hill process including as many new
estimators of the extremal index when $F$ is in the extremal domain. We
focus in this paper on its functional and uniform asymptotic law in the new
setting of weak convergence in the space of bounded real functions. The
results are next particularized for explicit examples of classes $\mathcal{F}
$.
\end{abstract}

\author{Gane Samb LO$^{*}$}
\author{El Hadj DEME$^{**}$}
\address{$^{*}$ LSTA, UPMC, France and LERSTAD, Universit\'e Gaston Berger
de Saint-Louis, SENEGAL\\
gane-samb.lo@ugb.edu.sn, ganesamblo@ufrsat.org}
\address{$^{**}$ LERSTAD, Universit\'e Gaston Berger de Saint-Louis, SENEGAL%
\\
ehdeme@ufrsat.org}
\maketitle

\section{Introduction}

%\Large

\noindent Let $X_{1},X_{2},...$ be a sequence of independent copies (s.i.c)
of a real random variable (r.v.) $X>1$ with d.f. $F(x)=\mathbb{P}(X\leq x)$.
We will be concerned in this paper with the functional asymptotic behaviour
of the sequence of stochastic processes 
\begin{equation}
T_{n}(f)=\sum_{j=1}^{k}f(j)\left( \log X_{n-j+1,n}-\log X_{n-j,n}\right) ,
\label{tf01}
\end{equation}
indexed by some classes $\mathcal{F}$ of functions $f: \mathbb{N}^{\ast
}= \mathbb{N}\backslash \{0\}\longmapsto \mathbb{R}_{+}$ and where $k=k(n)$ satisfies \begin{equation*}
1\leq k\leq n,k/n\rightarrow 0\text{ as }n\rightarrow \infty .
\end{equation*}
The main motivation of this study is to obtain very large classes of
estimators for the extreme value index when $F$ lies in the extremal domain,
all of them being margins of only one stochastic process. Indeed, for the
uniform function $f(j)=j,$ $k^{-1}T_{n}(f)$ is the famous Hill (\cite{hill})
estimator of such an index. Recently, a first step to functional forms of
the Hill estimator has been done in \cite{dioplo} in the form $k^{-\tau
}T_{n}(f)$ for $f(j)=j^{\tau }$ and respectively studied in \cite{dioplo}
for ($\tau>1/2)$ and in \cite{demedioplo} for $(0<\tau < 1/2)$ in finite
distributions. Groenboom and al. (\cite{glw}) also considered a thorough study of a family of
Kernel-type estimators of the extreme value index. However, they did not consider a stochastic processes view. There exists a very large number of estimators of the
extremal index. We may cite those of Cs\"{o}rg\H{o}-Deheuvels-Mason \cite%
{cdm}, De Haan-Resnick \cite{haanresnick}, Pickands \cite{pickands},
Deckkers, De Haan and Einmahl \cite{deh}, Hasofer and Wang \cite{hw}, etc.
But they all go back to the Hill's one.\newline

\bigskip \noindent However, the asymptotic theory for the estimators of the
extremal index are set for a finite number of them, in finite distributions
for whole the extremal domain $(-\infty \leq \gamma \leq +\infty)$. The
reader is referrred to following sample citations : \cite{cdm}, \cite%
{csorgo-mason}, \cite{deh}, \cite{dioplo}, \cite{haanresnick}, \cite{hall1}, 
\cite{hall2}, \cite{haeusler-teugels}, \cite{hw}, \cite{gslo}, \cite%
{pickands}, etc.\newline

\noindent Now the modern setting of functional weak convergence allows to
handle more complex estimators in form of stochastic processes, say $%
\{T_{n}(f),f\in \mathcal{F}\}$, such that for any $f\in \mathcal{F}$, there
exists a nonrandom sequence $a_{n}(f)$ such that $T_{n}/a_{n}(f)$ is an
estimator of the extremal index $\gamma$. Such processes may be called
stochastic processes of estimators of the extremal index.\newline

\noindent Here, we consider one of such processes, that is (\ref{tf01}). Our
aim is to derive their functional asymptotic normality when possible or
simply their asymptotic distribution for suitable classes. We will see that
for some classes, we have non Gaussian asymptotic behavior, which will be
entirely characterized. We shall mainly consider two classes of functions.
The first consists of those functions $f$ satisfying 
\begin{equation}
\text{ }A(2,f)=\sum_{j=1}^{\infty }f(j)^{2}j^{-2}<\infty ,  \tag{K1}
\end{equation}%
with the general notation $A(m,f)=\sum_{j=1}^{\infty }f(j)^{m}j^{-m}.$ The
second includes functions $f$ such that 
\begin{equation}
\limsup_{n\rightarrow +\infty }B(n,f)=0,  \tag{K2}
\end{equation}%
where $B(n,f)=\sigma _{n}(f)^{-1}\max \{f(j)j^{-1},1\leq j\leq k\}$ and $%
\sigma _{n}(f)$ is defined below in (\ref{defan}). Under these two
conditions, we will be able to find the asymptotic distributions of $%
T_{n}(f) $ for a fixed $f$, under usual and classical hypotheses of extreme
value Theory. But as to functional laws, we need uniform conditions. Define $%
\mathcal{F}_{1}$ the subclass of $\mathcal{F}$ such that 
\begin{equation}
0<\inf_{f\in \mathcal{F}_{1}}A(2,f)<\sup_{f\in \mathcal{F}%
_{1}}A(2,f)<+\infty,  \tag{KU1}
\end{equation}%
and \ $\mathcal{F}_{2}$ be a subclass of $\mathcal{F}$ such that 
\begin{equation}
\lim_{n\rightarrow \infty }\sup_{f\in F_{2}}B_{n}(f)=0,  \tag{KU2a}
\end{equation}%
and such that for any couple $(f_{1},f_{2})\in \mathcal{F}_{2},$%
\begin{equation}
\lim_{n\rightarrow \infty }\frac{1}{\sigma _{n}(f_{1})\sigma _{n}(f_{2})}%
\sum_{j=1}^{k}f_{1}(j)f_{2}(j)j^{-2}=\Gamma (f_{1},f_{2})  \tag{KU2b}
\end{equation}%
exists, where 
\begin{equation}
\sigma _{n}^{2}(f)=\sum_{j=1}^{k}f(j)^{2}j^{-2}\text{ and }%
a_{n}(f)=\sum_{j=1}^{k}f(j)j^{-1},.  \label{defan}
\end{equation}%
We will suppose at times that each $\mathcal{F}_{h}$ is totally bounded with
respect to some semimetric $\rho _{h}$.\newline

\noindent Our best achievement is the complete description of the weak
convergence of the sequence 
\begin{equation*}
\{T_{n}(f),f\in \mathcal{F}_{h}\},\text{ }h=1,2,
\end{equation*}%
in the spaces $\ell ^{\infty }(\mathcal{F}_{h})$ of bounded and real
functions defined on $\mathcal{F}_{h},$ in the light of the modern setting
of this theory. Further we provide real case studies with explicit classes
in application of the general results.\newline

\bigskip \noindent This approach yields a very great number of estimators of
the tail distribution $1-F$ in the extreme value domain. But, this paper
will essentially focus on the functional and uniform laws of the process
described above. Including in the present work, for example, second and
third order conditions as it is fashion now, and considering datadriven
applications or simulations studies would extremely extend the report. These
questions are to be considering in subsequent papers.

\bigskip

\noindent This paper will use technical results of extreme value Theory. So
we will summarize some basics of this theory in the section \ref{sec2}. In
this section, we introduce basic notation and usual representation of
distribution functions lying in the extremal domain, to be used in all the
remainder of the paper. Section \ref{sec3} is devoted to pointwise limit
distributions of $T_{n}(f),$ while our general functional results are stated
and established in Section \ref{sec4}. In Section \ref{sec5}, we study some
particular cases, especially the family $\{f(j)=j^{\tau },\tau >0\}.$ The
section \ref{sec6} is devoted of the tools of the paper. In this latter, we
state two lemmas, namely Lemmas 1 and 2, which are key tools in the earlier
proofs.

\section{Some basics of Extreme Value Theory}

\label{sec2}

\noindent The reader is referred to de Haan (\cite{dehaan} and \cite{dehaan1}%
), Resnick (\cite{resnick}), Galambos (\cite{galambos}) and Beirlant,
Goegebeur and Teugels (\cite{bgt}) for a modern and large account of the
extreme value theory. A distribution function F is said to be attracted to a
non degenerated $M$ iff the maximum $X_{n,n}=\max \left(
X_{1},...X_{n}\right) ,$ when appropriately centred and normalized \ by two
sequences of real numbers $\left( a_{n}>0\right) _{n\geq 0}$ and $\left(
b_{n}\right) _{n\geq 0}$, converges to M, in the sense that

\begin{equation*}
\lim_{n\rightarrow +\infty }P\left( X_{n,n}\leq a_{n}\text{ }x+b_{n}\right)
\end{equation*}

\begin{equation}
=\lim_{n\rightarrow +\infty }F^{n}\left( a_{n}x+b_{n}\right) =M(x),
\label{dl05}
\end{equation}
for continuity points x of M.\newline

\noindent If (\ref{dl05}) holds, it is said that $F$ is attracted to $M$ or $%
F$ belongs to the domain of attraction of $M$, written $F\in D(M).$ It is
well-kwown that the three possible nondegenerate limits in (\ref{dl05}),
called extremal d.f., are the following \newline

\noindent The Gumbel $d.f.$ 
\begin{equation}
\Lambda (x)=\exp (-\exp (-x)),\text{ }x\in \mathbb{R},  \label{dl05a}
\end{equation}%
or the Fr\'{e}chet $d.f.$ of parameter $\alpha >0,$

\begin{equation}
\phi _{\gamma }(x)=\exp (-x^{-\alpha })\mathbb{I}_{\left[ 0,+\infty \right[
}(x),\text{ }x\in \mathbb{R}\   \label{dl05b}
\end{equation}%
or the Weibull $d.f.$ of parameter $\alpha >0$

\begin{equation}
\psi _{\gamma }(x)=\exp (-(x)^{-\alpha })\mathbb{I}_{\left] -\infty ,0\right]
}(x)+(1-1_{\left] -\infty ,0\right] }(x)),\ \ x\in \mathbb{R},\ 
\label{dl05c}
\end{equation}%
where $\mathbb{I}_{A}$ denotes the indicator function of the set A. Now put $%
D(\phi )=\cup _{\alpha >0}D(\phi _{\gamma }),$ $D(\psi )=\cup _{\alpha
>0}D(\psi _{\gamma }),$ and $\Gamma =D(\phi )\cup D(\psi )\cup D(\Lambda )$.%
\newline

\bigskip

\noindent In fact the limiting distribution function $M$ is defined by an
equivalence class of the binary relation $\mathcal{R}$ on the set of d.f's $%
\mathcal{D}$ in $\mathbb{R}$ defined as follows : 
\begin{equation*}
\forall (M_{1},M_{2})\in \mathcal{D}^{2},(M_{1}\text{ }\mathcal{R}\text{ }%
M_{2})\Leftrightarrow \exists (a,b)\in \mathbb{R}_{+}\backslash \{0\}\times 
\mathbb{R},\forall (x\in \mathbb{R}),
\end{equation*}
\begin{equation*}
M_{2}(x)=M_{1}(ax+b).
\end{equation*}
One easily checks that if $F^{n}\left( a_{n}x+b_{n}\right) \rightarrow
M_{1}(x),$ then $F^{n}\left( c_{n}x+d_{n}\right) \rightarrow
M_{1}(ax+b)=M_{2}(x)$ whenever 
\begin{equation}
a_{n}/d_{n}\rightarrow a\text{ and }(b_{n}-d_{n})/c_{n}\rightarrow b\text{
as }n\rightarrow \infty .  \label{dl05f}
\end{equation}

\noindent Theses facts allow to parameterize the class of extremal
distribution functions. For this purpose, suppose that (\ref{dl05}) holds
for the three d.f.'s given in (\ref{dl05a}), (\ref{dl05b}) and (\ref{dl05c}%
). We may take sequences $(a_{n}>0)_{n\geq 1}$ and $(b_{n})_{n\geq 1}$ such
that the limits in (\ref{dl05f}) are $a=\gamma =1/\alpha $ and $b=1$ (in the
case of Fr\'{e}chet extremal domain), and $a=-\gamma =-1/\alpha $ and $b=-1$
(in the case of Weibull extremal domain). Finally, one may interprets $%
(1+\gamma x)^{-1/\gamma }=exp(-x)$ for $\gamma =0$ (in the case of Gumbel
extremal domain). This leads to the following parameterized extremal
distribution function 
\begin{equation}
G_{\gamma }(x)=\exp (-(1+\gamma x)^{-1/\gamma }),\text{ }1+\gamma x\geq 0,
\label{dl05d}
\end{equation}%
called the Generalized Pareto Distribution (GPD) of parameter $\gamma \in 
\mathbb{R}$.

\bigskip

\noindent Now we give the usual representations of $df^{\prime }s$ lying in
the extremal domain in terms of the quantile function of G(x)=F(e$%
^{x}),x\geq 1,$ that is $G^{-1}(1-u)=\log F^{-1}(1-u),0\leq u\leq 1.$

\begin{theorem}
\label{theo1} We have :

\begin{enumerate}
\item Karamata's representation (KARARE)

\noindent (a) If $F\in D(\phi _{1/\gamma }),$ $\gamma >0$, then 
\begin{equation}
G^{-1}(1-u)=\log c+\log (1+p(u))-\gamma \log u+(\int_{u}^{1}b(t)t^{-1}dt),%
\text{ }0<u<1,  \label{rep1}
\end{equation}
where $\sup (\left| p(u)\right| ,\left| b(u)\right| )\rightarrow 0$ as $%
u\rightarrow 0$ and c is a positive constant and $G^{-1}(1-u)=\inf
\{x,G(x)\geq u\},$ $0\leq u\leq 1,$ is the generalized inverse of $G$ with $%
G^{-1}(0)=G^{-1}(0+)$.\newline
\newline
\noindent (b) If $F\in D(\psi _{1/\gamma }),$ $\gamma >0$, then $%
y_{0}(G)=\sup \{x,$ $G(x)<1\}<+\infty $ and 
\begin{equation}
y_{0}-G^{-1}(1-u)=c(1+p(u))u^{\gamma }\exp (\int_{u}^{1}b(t)t^{-1}dt),\text{ 
}0<u<1,  \label{rep2}
\end{equation}
where $c$, $p(\cdot )$ and $b(\cdot )$ are as in (\ref{rep1})

\item Representation of de Haan (Theorem 2.4.1 in \cite{dehaan}),

\noindent If $G\in D(\Lambda )$, then 
\begin{equation}
G^{-1}(1-u)=d-s(u)+\int_{u}^{1}s(t)t^{-1}dt,\text{ }0<u<1,  \label{rep3a}
\end{equation}%
\noindent where d is a constant and $s(\cdot )$  admits this KARARE : 
\begin{equation}
s(u)=c(1+p(u))\exp (\int_{u}^{1}b(t)t^{-1}dt),\text{ }0<u<1,  \label{rep3b}
\end{equation}%
\noindent $c$, $p(\cdot )$ and $b(\cdot )$ being defined as in (\ref{rep1}).
\end{enumerate}
\end{theorem}

\bigskip

\bigskip 

\noindent We restrict ourselves ourselves here to the cases $F \in D(\Gamma) \cup D(\Phi_{1/\gamma})$, $\gamma >0$,  since the case $F\in D(\Psi_{1/\gamma})$, $\gamma >0$, may be studied through the transform  $F(x_{0}(F)-1/\dot{})) \in D(\Phi_{1/\gamma})$ for estimating $\gamma$. This leads to replacle $X_{n-j+1,n}$ by $x_{0}(F)-1/1/X_{j,n}$ in (\ref{tf01}). However, a direct investigation of (\ref{tf01}) for $F\in D(\Psi_{1/\gamma})$, $\gamma >0$ is possible. This requires the theory of sums of dependent random variables while this paper uses sums if independant random varaibles, as it will seen shortly. We consequently  consider a special handling of this cas in a dinstinct paper.\\

\noindent Finally, we shall also use the uniform representation of $%
Y_{1}=\log X_{1},Y_{2}=\log X_{2},...$ by $%
G^{-1}(1-U_{1}),G^{-1}(1-U_{2}),...$ where $U_{1},U_{2},...$ are independent
and uniform random variables on $(0,1)$ and where $G$ is the $d.f.$ of $Y$,
in the sense of equality in distribution (denoted by $=_{d})$%
\begin{equation*}
\left\{ Y_{j},j\geq 1\}=_{d}\{G^{-1}(1-U_{j}),j\geq 1\right\} ,
\end{equation*}%
and hence 
\begin{equation}
\{\left\{ Y_{1,n},Y_{2,n},...Y_{n,n}\right\} ,n\geq 1\}  \label{repre}
\end{equation}

\begin{equation*}
=_{d}\left\{
\{G^{-1}(1-U_{n,n}),G^{-1}(1-U_{n-1,n}),...,G^{-1}(1-U_{1,n})\},n \geq
1\right\}.
\end{equation*}

\noindent In connexion with this, we shall use the following Malmquist
representation (see (\cite{shwell}), p. 336) :

\begin{equation}
\{\log (\frac{U_{j+1,n}}{U_{j,n}})^{j},j=1,...,n\}=_{d}\{E_{1},...,E_{n}\},
\label{malm}
\end{equation}
where $E_{1},...,E_{n}$ are independent standard exponential random
variables.

\section{Pointwise and Finite-distribution Laws of $T_{n}(f)$}

\label{sec3}

\noindent Let us begin to introduce these conditions on the distribution
function $G$, through the functions $p$ and $b$ in the representations (\ref%
{rep1}), (\ref{rep2}), (\ref{rep3a}) and (\ref{rep3b}). First, define for $%
\lambda >1$, 
\begin{equation*}
0\leq g_{1,n}(p,\lambda )=\sup_{0\leq u\leq \lambda k/n}\left\vert
p(u)\right\vert ,
\end{equation*}%
\begin{equation*}
g_{2,n}(b,\lambda )=\sup_{0\leq u\leq \lambda k/n}\left\vert b(u)\right\vert
,
\end{equation*}%
and 
\begin{equation*}
d_{n}(p,b,\lambda )=\max (g_{1,n}(p,\lambda ),g_{2,n}(b,\lambda )\log k).
\end{equation*}%
We will need the following conditions for some $\lambda >1$ :

\begin{equation}
g_{1,n}(f,\lambda )(\sigma _{n}(f))^{-1}\sum_{j=1}^{k}f(j)\rightarrow 0,%
\text{ }as\text{ }n\rightarrow \infty ,  \tag{C1}
\end{equation}%
\begin{equation}
g_{2,n}(b,\lambda )(\sigma _{n}(\tau )k^{\tau
})^{-1}\sum_{j=1}^{k}f(j)\rightarrow 0\text{ }as\text{ }n\rightarrow \infty 
\tag{C2}
\end{equation}%
and 
\begin{equation}
\sup_{f\in \mathcal{F}_{h}}d_{n}(p,b,\lambda )(\sigma _{n}(f)k^{\tau
})^{-1}\sum_{j=1}^{k}f(j)\rightarrow 0\text{ }as\text{ }n\rightarrow \infty .
\tag{C3}
\end{equation}%
From now on, all the limits are meant as $as$ $n\rightarrow \infty $ unless
the contrary is specified. We are able to state :

\begin{theorem}
\label{theo2} \bigskip Let $F\in D(\Lambda ).$ If (C3) holds, then 
\begin{equation*}
(s(k/n)\sigma _{n}(f))^{-1}(T_{n}(f)-a_{n}(f)s(k/n))\rightarrow \mathcal{N}%
(0,1)
\end{equation*}%
when $(K2)$ holds and 
\begin{equation*}
(s(k/n)\sigma _{n}(f))^{-1}(T_{n}(f)-a_{n}(f)s(k/n))\rightarrow \mathcal{L}%
(f),
\end{equation*}%
when $(K1)$ is satisfied, and where 
\begin{equation*}
\mathcal{L}(f)=A(2,f)^{-1/2}\sum_{j=1}^{\infty }f(j)j^{-1}(E_{j}-1).
\end{equation*}%
Let $F\in D(\phi _{1/\gamma })$. If (C1) and (C2) hold, then 
\begin{equation*}
(a_{n}(f)/\sigma _{n}(f))(T_{n}(f)/a_{n}(f)-\gamma )\rightarrow \mathcal{N}%
(0,\gamma ^{2})
\end{equation*}%
under $(K2)$ and 
\begin{equation*}
(a_{n}(f)/\sigma _{n}(f))(T_{n}(f)/a_{n}(f)-\gamma )\rightarrow \gamma ^{-1}%
\mathcal{L}(f)
\end{equation*}%
under $(K1)$.
\end{theorem}

\begin{proof}
Let us use the representation (\ref{repre}). We thus have, for any $n\geq 1,$%
\begin{equation*}
\left\{ \log X_{n-j+1,n}=Y_{n-j+1,n},1\leq j\leq n\right\} =_{d}\left\{
G^{-1}\{1-U_{j,n}),1\leq j\leq n\right\} .
\end{equation*}%
First, let $F\in D(\Lambda ).$ By (\ref{rep2}), we get 
\begin{equation*}
T_{n}(f)=\sum_{j=1}^{k}f(j)(s(U_{j,n})-s(U_{j+1,n}))+\sum_{j=1}^{k}f(j)%
\int_{U_{j,n}}^{U_{j+1,n}}s(t)/t\text{ }dt
\end{equation*}

\begin{equation*}
\equiv S_{n}(1)+S_{n}(2).
\end{equation*}
Using (\ref{rep3b}), we have for $U_{1,n}\leq v,u\leq U_{k,n},$%
\begin{equation*}
s(u)/s(v)=(1+p(u))/(1+p(v))\exp (-\int_{U_{1,n}}^{U_{k+1,n}}t^{-1}b(t)dt).
\end{equation*}
Putting 
\begin{equation}
g_{1,n,0}(p)=\sup \{\left| p(u)\right| ,0\leq u\leq U_{k+1,n}\}\text{ and }%
g_{2,n,0}(p)=\sup \{\left| b(u)\right| ,0\leq u\leq U_{k+1,n}\},
\label{preuve3}
\end{equation}
we get, since $\log (U_{k+1,n}/U_{1,n})=O_{p}(\log k)$ as $n\rightarrow
\infty ,$%
\begin{equation*}
s(u)/s(v)=(1+O(g_{1,n,0}))\exp (-O_{p}(g_{2,n,0}\log k)).
\end{equation*}
This implies 
\begin{equation}
\sup_{U_{1,n}\leq u,v\leq U_{k,n}}\left| s(u)/s(v)-1\right| =O_{p}(\max
(g_{1,n,0},g_{2,n,0}\log k))  \label{preuve2}
\end{equation}
as $n\rightarrow \infty $ and 
\begin{equation}
\sup_{U_{1,n}\leq u,v\leq U_{k,n}}\left| \frac{s(u)-s(v)}{s(k/n)}\right|
=O_{p}(\max (g_{1,n,0},g_{2,n,0}\log k)).  \label{preuve0}
\end{equation}
Since $nk^{-1}U_{k+1,n}\rightarrow 1$ $a.s.$ as $n\rightarrow \infty ,$ we
may find for any $\varepsilon >0$\ and for any $\lambda >1,$ an integer $%
N_{0}$ such that for any $n\geq N_{0},$%
\begin{equation}
\mathbb{P}(g_{1,n,0}\leq g_{1,n}(p,\lambda ),\text{ }g_{2,n,0}\leq
g_{2,n}(b,\lambda )) \geq 1-\varepsilon .  \label{preuve1}
\end{equation}
Hence $(C3)$ implies 
\begin{equation*}
\frac{S_{n}(1)}{\sigma _{n}(f)s(k/n)}\leq d_{n}(p,b,\lambda )\text{ }(\sigma
_{n}(f))^{-1}\sum_{j=1}^{k}f(j)\rightarrow _{P}0,
\end{equation*}

\noindent where $d_{n}(p,b,\lambda )=\max (g_{1,n},g_{2,n}\log k).$ Next

\begin{equation*}
\frac{S_{n}(2)}{\sigma _{n}(f)s(k/n)}=\sigma
_{n}(f)^{-1}\sum_{j=1}^{k}f(j)\int_{U_{j,n}}^{U_{j+1,n}}\left\{
s(t)/s(k/n)\right\} /t\text{ }dt
\end{equation*}
\begin{equation*}
=\sigma _{n}(f)^{-1}\sum_{j=1}^{k}f(j)\int_{U_{j,n}}^{U_{j+1,n}}t^{-1}\text{ 
}dt
\end{equation*}
\begin{equation*}
+\sigma _{n}(f)^{-1}\sum_{j=1}^{k}f(j)\int_{U_{j,n}}^{U_{j+1,n}}\left\{
s(t)/s(k/n)-1\right\} /t\text{ }dt=S_{n}(2,1)+S_{n}(2,2).
\end{equation*}
We have, by (\ref{preuve2}) and the Malmquist representation (\ref{malm}), 
\begin{equation*}
\left| S_{n}(2,2)\right| \leq O_{p}(1)d_{n}(p,b,\lambda )\times \sigma
_{n}(f)^{-1}\sum_{j=1}^{k}f(j)j^{-1}E_{j}\leq O_{p}(1)\text{ }\times
\end{equation*}
\begin{equation*}
\left\{ d_{n}(p,b,\lambda )\times \sigma
_{n}(f)^{-1}\sum_{j=1}^{k}f(j)j^{-1}(E_{j}-1)+d_{n}(p,b,\lambda )\times
\sigma _{n}(f)^{-1}\sum_{j=1}^{k}f(j)j^{-1}\right\} .
\end{equation*}
The first term tends to zero since $\sigma
_{n}(f)^{-1}\sum_{j=1}^{k}f(j)j^{-1}(E_{j}-1)$ converges in distribution to
a finite random variable by Lemma \ref{lemmatool} in Section \ref%
{sectiontool} and $d_{n}(p,b,\lambda )\rightarrow _{P}0$ $by$ in probability
(\ref{preuve2}). The second also tends to zero by $(C3).$ Finally, by the
Malmquist representation (\ref{malm}), one arrives to 
\begin{equation*}
S_{n}(2,1)=\sigma _{n}(f)^{-1}\sum_{j=1}^{k}f(j)j^{-1}E_{j}.
\end{equation*}

\noindent And this leads to 
\begin{equation*}
S_{n}(2,1)-\left\{ a_{n}(f)/\sigma _{n}(f)\right\} =\sigma _{n}(\tau
)^{-1}\sum_{j=1}^{k}j^{\tau -1}(E_{j}-1),
\end{equation*}%
which converges in distribution to a $\mathcal{N}(0,1)$ random variable
under (K2) and to $\mathcal{L}(f)$ under (K1) by Lemma \ref{lemmatool} in
Section \ref{sectiontool}. By summerizing all these facts, we have proved
that 
\begin{equation*}
(\sigma _{n}(f)s(k/n))^{-1}(T_{n}(f)-a_{n}(f)s(k/n))
\end{equation*}%
converges in distribution to a $\mathcal{N}(0,1)$ random variable under $%
(K2) $ and to $\mathcal{L}(f)$ under (K1)$.$

\noindent Now let $F\in D(\phi _{1/\gamma}),$ we have by (\ref{rep1}) and
the usual representations, 
\begin{equation*}
T_{n}(f)=\sum_{j=1}^{k}f(j)\left\{ \log (1+p(U_{j+1,n}))-\log
(1+p(U_{j,n}))\right\}
\end{equation*}
\begin{equation*}
+\gamma \sum_{j=1}^{k}f(j)\log
(U_{j+1,n}/U_{j,n})+\sum_{j=1}^{k}f(j)\int_{U_{j,n}}^{U_{j+1,n}}b(t)/t\text{ 
}dt
\end{equation*}

\begin{equation*}
\equiv S_{n}(1)+S_{n}(2)+S_{n}(3).
\end{equation*}
We have, for large values of $k$, 
\begin{equation*}
\left| S_{n}(1)/\sigma _{n}(f)\right| \leq 2g_{1,n,0}(f)(\sigma
_{n}(f))^{-1}\sum_{j=1}^{k}f(j),
\end{equation*}
where $g_{1,n,0\text{ }}$is defined in (\ref{preuve3}), which tends to zero
in probability by $(C1)$ and (\ref{preuve2}). Next 
\begin{equation*}
\left| S_{n}(3)/\sigma _{n}(\tau )\right| \leq g_{2,n,0}(b)(\sigma
_{n}(f))^{-1}\sum_{j=1}^{k}f(j)\log (U_{j+1,n}/U_{j,n})
\end{equation*}
\begin{equation*}
=g_{2,n,0}(b)\sigma _{n}(f)^{-1}\sum_{j=1}^{k} f(j)j^{-1}
(E_{j}-1)+g_{2,n,0}(b)\sigma _{n}(f)^{-1}\sum_{j=1}^{k}f(j)j^{-1},
\end{equation*}
where $g_{2,n,0\text{ }}$defined in (\ref{preuve3}). Then $S_{n}(3)/\sigma
_{n}(f)\rightarrow 0$ by $(C3)$ and Lemma \ref{lemmatool} and the methods
described above. Finally, always by Lemma \ref{lemmatool}, 
\begin{equation*}
\left\{ (S_{n}(3)-\gamma a_{n}(f)\right\} /\sigma _{n}(f)=\gamma \sigma
_{n}(f)^{-1}\sum_{j=1}^{k}f(j)j^{-1}(E_{j}-1)
\end{equation*}
and this converges in distribution to a $\mathcal{N}(0,\gamma ^{2})$ random
variable under (K2) and to $\gamma L(f)$ under (K1).
\end{proof}

\bigskip

\noindent From these proofs, we get two intermediate results towards the
functional laws. The first concerns the asymptotic law in finite
distributions.

\begin{corollary}
\label{cor1} Suppose that the hypotheses (C1), (C2) and (C3) hold and for
any $(f_{1},f_{2})\in \mathcal{F}_{1}^{2}\cup \mathcal{F}_{2}^{2}$, 
\begin{equation*}
\lim_{n\rightarrow \infty }\frac{1}{\sigma _{n}(f_{1})\sigma _{n}(f_{2})}%
\sum_{j=1}^{k}f_{1}(j)f_{2}(j)j^{-2}=\Gamma (f_{1},f_{2})\text{ }exists.
\end{equation*}%
Then the finite-distributions of $\{(s(k/n)\sigma
_{n}(f))^{-1}(T_{n}(f)-a_{n}(f)s(k/n)),f\in \mathcal{F}_{1}\}$ weakly
converge to those of the process $\mathcal{L}$, for $F\in D(\Lambda )$ and
the finite-distributions of $\{\sigma _{n}(f)^{-1}(T_{n}(f)-a_{n}(f),f\in 
\mathcal{F}_{1}\}$ weakly converge to those of the process $\gamma \mathcal{L%
}$, for $F\in D(\varphi _{1/\gamma })$.

\noindent And the finite-distributions of $\{(s(k/n)\sigma
_{n}(f))^{-1}(T_{n}(f)-a_{n}(f)s(k/n)),f\in \mathcal{F}_{2}\}$ weakly
converge to those of a Gausian process $\mathbb{G}$\ of covariance function 
\begin{equation*}
\Gamma (f_{1},f_{2})=\lim_{n\rightarrow \infty }\frac{1}{\sigma
_{n}(f_{1})\sigma _{n}(f_{2})}\sum_{j=1}^{k}f_{1}(j)f_{2}(j)j^{-2}
\end{equation*}%
and, for $F\in D(\Lambda )$, the finite-distributions of $\{\sigma
_{n}(f)^{-1}(T_{n}(f)-a_{n}(f),f\in \mathcal{F}_{2}\}$ weakly converge to
those of $\gamma \mathbb{G}$, for $F\in D(\varphi _{1/\gamma }).$
\end{corollary}

\begin{proof}
Put $V_{n}(0,f)=(s(k/n)\sigma _{n}(f))^{-1}(T_{n}(f)-a_{n}(f)s(k/n))$ and $%
V_{n}(1,f)=\sigma _{n}(f)^{-1}(T_{n}(f)-a_{n}(f))$ for a fixed $f\in 
\mathcal{F}_{h},h=1,2.$ For a finite family $(f_{1},f_{2},...,f_{s})\in f\in 
\mathcal{F}_{h}^{s},h=1,2,$ $0<s\in \mathbb{N}^{\ast },$ we have by the
proof of Theorem \ref{theo1}, for each $1\leq i\leq s,$ 
\begin{equation}
V_{n}(0,f_{i})=V_{n}^{\ast }(f_{i})+o_{P}(1)  \label{fd01}
\end{equation}
for $F\in D(\Lambda )$ and%
\begin{equation}
\text{ }V_{n}(1,f_{i})=\gamma V_{n}^{\ast }(f_{i})+o_{P}(1)  \label{fd02}
\end{equation}%
for $F\in D(\varphi _{1/\gamma }).$ From now, we conclude by applying Lemma %
\ref{lemma2} which establishes the asymptotic laws of \ $(V_{n}^{\ast
}(f_{1}),...,V_{n}^{\ast }(f_{s}))$ under the assumptions of the Lemma.
\end{proof}

\bigskip

\noindent This corollary makes a good transition towards the functional law.
In order to state a further step, we need the following uniform conditions
on the distribution function $F$. Define for some $\lambda >1,$

\begin{equation}
\sup_{f\in \mathcal{F}_{h}}g_{1,n}(f,\lambda )(\sigma
_{n}(f))^{-1}\sum_{j=1}^{k}f(j)\rightarrow 0,  \tag{CU1}
\end{equation}
\begin{equation}
\sup_{f\in \mathcal{F}_{h}}g_{2,n}(b,\lambda )\sigma
_{n}(f)^{-1}\sum_{j=1}^{k}f(j)  \tag{CU2}
\end{equation}
and 
\begin{equation}
\sup_{f\in \mathcal{F}_{h}}d_{n}(p,b,\lambda )\sigma
_{n}(f)^{-1}\sum_{j=1}^{k}f(j)\rightarrow 0.  \tag{CU3}
\end{equation}

\noindent These conditions are set so that the $o_{P}(1)$ in (\ref{fd01})
and \ref{fd02}) hold uniformly in our classes. We thus begin to state this :

\begin{corollary}
\label{cor2} Assume that the uniform conditions (KU1), (KU2a), (KU2b),
(CU1), (CU2) and (CU3) hold - when appropriate - in Theorem \ref{theo2}. Put 
$V_{n}^{\ast }(f)=\sigma _{n}(f)^{-1}(V_{n}(f)-a_{n}(f))$. Finally let $%
\mathcal{F}_{a}$ be a nonvoid family of functions satisfying $(KU1)$ or, a
nonvoid family of functions satisfying $(KU2a-b)$. Suppose that $%
\{V_{n}^{\ast }(f),f\in \mathcal{F}_{a}\}$ weakly converges in $\ell
^{\infty }(\mathcal{F}_{a})$. Then, uniformly in $f\in \mathcal{F}_{a}$,

\begin{equation*}
(s(k/n)\sigma _{n}(f))^{-1}(T_{n}(f)-a_{n}(f)s(k/n))=V_{n}^{\ast
}(f)+o_{P}^{\ast }(1)
\end{equation*}

\noindent for $F \in D(\Lambda)$ and

\begin{equation*}
(a_{n}(f)/\sigma _{n}(f)) (T_{n}(f)/a_{n} - \gamma)= \gamma V_{n}^{\ast
}(f)+o_{P}^{\ast }(1)
\end{equation*}

\noindent for $F \in D(\phi_{1/\gamma})$.
\end{corollary}

\begin{proof}
\bigskip Put 
\begin{equation*}
V_{n}(0,f)=(s(k/n)\sigma_{n}(f))^{-1}(T_{n}(f)-a_{n}(f)s(k/n))
\end{equation*}
\noindent and 
\begin{equation*}
V_{n}(1,f)=(a_{n}(f)/\sigma_{n}(f))(T_{n}(f)/a_{n}-\gamma ). 
\end{equation*}

\noindent When the uniformity hypotheses (KU1) or,(KU2a) and (KU2b) hold, we
surely have 
\begin{equation*}
V_{n}(0,f)=V_{n}^{\ast }(f)(1+o_{P}^{\ast }(1))+o_{P}^{\ast }(1)\text{ and }%
V_{n}(1,f)=\gamma V_{n}^{\ast }(f)(1+o_{P}^{\ast }(1))+o_{P}^{\ast }(1),
\end{equation*}%
uniformly in $f\in \mathcal{F}_{h}.$\ Now let $\mathcal{F}_{a}$ a subset of $%
\mathcal{F}_{1}$ such that $\{V_{n}^{\ast }(f),f\in \mathcal{F}_{a}\}$
weakly converges, say to $\mathbb{G}$ in $\ell ^{\infty }(\mathcal{F}_{a})$.
Then $\left\Vert \mathbb{G}\right\Vert _{\mathcal{F}_{a}}^{\ast }<\infty $.
Since, by the continuity theorem, $\left\Vert V_{n}^{\ast }\right\Vert _{%
\mathcal{F}_{a}}\leadsto \left\Vert \mathbb{G}\right\Vert _{\mathcal{F}%
_{a}}^{\ast },$ we get 
\begin{equation}
V_{n}(0,f)=V_{n}^{\ast }(f)+o_{P}^{\ast }(1),  \label{udf01}
\end{equation}%
uniformly in $f\in \mathcal{F}_{a}$ for $F\in D(\Lambda ).$ The other cases
are proved similarly.
\end{proof}

\section{The functional law of $T_{n}(f)$}

\label{sec4}

We shall use here (\ref{udf01}). Recall and denote%
\begin{equation*}
V_{n}^{\ast }(f)=\sigma _{n}(f)^{-1}(V_{n}(f)-a_{n}(f))=\sum_{j=1}^{k}\frac{%
f(j)j^{-1}}{\sigma _{n}(f)}(E_{j}-1)=:\sum_{j=1}^{k}Z_{j,n}(f).
\end{equation*}

\noindent Then, our main tools for handling these stochastic processes are
Theorem 2.11.1 and Theorem 2.11.9 of (\cite{vaart}) on uniform laws of sums
of independant stochastic processes. We will then need the basic frame of
these theorems.\newline

\noindent Next, we shall possibly consider sub-families $\mathcal{F}$\ of of 
$\mathcal{F}_{h}$ $(h=1,2)$\ formed by functions $f$ : $\mathbb{N}^{\ast
}\longmapsto \mathbb{R}_{+}^{\ast }$ satisfying this mesurability
assumption, that is for each $h=1,2,$ for each $\delta >0$ and for each $%
(e_{1},...,e_{n})\in \{-1,0,1\}^{n},$ for each $p=1,2$,%
\begin{equation}
\sup_{(f_{1},f_{2})\in \mathcal{F}_{h},\rho (f_{1},f_{2})\leq \delta
}\sum_{j=1}^{k}e_{j}\left\vert \sigma _{n}^{-1}(f_{1})f_{1}(j)j^{-1}-\sigma
_{n}^{-1}(f_{2})f_{2}(j)j^{-1}\right\vert ^{p}E_{j}^{p},  \tag{MES}
\end{equation}%
is measurable, where $\rho $ is a semimetric space on $\mathcal{F}$.
Precisely, we introduce the two classes. Let subfamilies $\mathcal{F}_{h,0}$
of $\mathcal{F}_{h}$ $(h=1,2)$ such that each of them is equiped with a
semimetric $\rho _{h}$ such that $(F_{h,0},\rho _{h})$ is totally bounded,
and that the mesurability of (MES) holds. We may also need the random
semimetric 
\begin{equation*}
d_{n}^{2}(f_{1},f_{2})=\sum_{j=1}^{k}(\sigma
_{n}^{-1}(f_{1})f_{1}(j)j^{-1}-\sigma
_{n}^{-1}(f_{2})f_{2}(j)j^{-1})^{2}(E_{j}-1)^{2}.
\end{equation*}%
Now suppose that $(\mathcal{F}_{h,0},\left\Vert {}\right\Vert )$ be a normed
space verifying the Riesz property. Let us define, as in \cite{vaart} (p.
211), the bracketing number $N_{\left[ {}\right] }(\varepsilon ,\mathcal{F}%
_{h,0},L_{2}^{n})$\ as the minimal number of sets $N_{\varepsilon }$ in a
partition $\mathcal{F}_{h,0}=\bigcup\limits_{j=1}^{N_{\varepsilon }}\mathcal{%
F}_{\varepsilon j}^{n}$ of the index set into sets $\mathcal{F}_{\varepsilon
j}^{n}$ such that, for every partitioning set $\mathcal{F}_{\varepsilon
j}^{n},$ 
\begin{equation*}
\sum_{i=1}^{k}\mathbb{E}^{\ast }\sup_{f,g\in F_{\varepsilon
j}^{n}}\left\vert Z_{ni}(f_{1})-Z_{ni}(f_{2})\right\vert ^{2}\leq
\varepsilon ^{2}.
\end{equation*}%
We have our first version of the functional laws of $T_{n}(f).$

\begin{theorem}
\label{theo3} Suppose that for each $h=1,2,$ we have 
\begin{equation}
\sup_{(f_{1},f_{2})\in \mathcal{F}_{h,0}^{2},\rho _{h}(f_{1},f_{2})\leq
\delta _{n}}\sum_{j=1}^{k(n)}(f_{1}(j)/(j\sigma _{n}(f))-f_{2}(j)/(j\sigma
_{n}(f)))^{2}\rightarrow 0  \tag{L1}
\end{equation}%
as $\delta _{n}\downarrow 0$ as $n\uparrow +\infty $, and 
\begin{equation}
\int_{0}^{\delta _{n}}\sqrt{logN_{[]}(\epsilon ,\mathcal{F}_{h},d_{n})}%
d\epsilon \rightarrow 0,  \tag{L2}
\end{equation}

\noindent as $n\uparrow \infty ,$ where $N_{\left[ {}\right] }(\epsilon ,%
\mathcal{F}_{h,0},d_{n})$ is the $\epsilon $-entropy number of $\mathcal{F}%
_{h,0}$ with respect to the semi-metric $d_{n}$, that is the minimal number
of $d_{n}$-balls of radius at most $\epsilon $ needed to cover $\mathcal{F}%
_{h}.$

\noindent Let $F\in D(\Lambda )$ and suppose that (CU3) holds. Then 
\begin{equation*}
\left\{ \left( s(k/n)\sigma _{n}(f)\right)
^{-1}(T_{n}(f)-s(k/n)a_{n}(f)),f\in \mathcal{F}_{1}\right\}
\end{equation*}%
converges to a Gaussian process in $\ell ^{\infty }(\mathcal{F}_{2,0})$ with
covariance function 
\begin{equation*}
\Gamma (f_{1},f_{2})=\lim_{n\rightarrow \infty }\sum_{j=1}^{k(n)}\left\{
\sigma _{n}(f_{1})\sigma _{n}(f_{2})\right\} ^{-1}f_{1}(j)f_{2}(j)j^{-2}\leq
1.
\end{equation*}

\noindent And $\left\{ \left( s(k/n)\sigma _{n}(f)\right)
^{-1}(T_{n}(f)-s(k/n)a_{n}(f)),f\in \mathcal{F}_{1,0}\right\} $ converges to
a stochastic process \{$\mathcal{L}(f),$ $f\in \mathcal{F}_{1,0}\}$ in $\ell
^{\infty }(\mathcal{F}_{1,0})$ with covariance function $\Gamma
(f_{1},f_{2}).$ The finite distributions of $(\mathcal{L}(f_{1}),...,%
\mathcal{L}(f_{S}))$ are characterized by the generating moments function 
\begin{equation}
(t_{1},...,t_{S})\mapsto \prod_{j=1}^{+\infty }\exp
(\sum_{s=1}^{S}t_{s}f_{s}(j)j^{-1})(1-\sum_{s=1}^{S}t_{s}f_{s}(j)j^{-1}),
\label{df}
\end{equation}%
for $\left\vert t_{s}\right\vert \leq 1/S,$ $s=1,...,S$.\newline

\noindent Let $F\in D(\phi _{1/\gamma })$ and suppose that (CU1) and (CU2)
hold. Then 
\begin{equation*}
\left\{ (a_{n}(f)/\sigma _{n}(f))(T_{n}(f)/a_{n}-\gamma ),f\in \mathcal{F}%
_{2,0}\right\} 
\end{equation*}

\noindent converges to a Gaussian process in $\ell ^{\infty }(\mathcal{F}%
_{2,0})$ with covariance function $\Gamma $.\newline

\noindent And $\left\{ (a_{n}(f)/\sigma _{n}(f))(T_{n}(f)/a_{n}-\gamma
)/\gamma ,f\in \mathcal{F}_{1,0}\right\} $ converges to a stochastic process
in $\ell ^{\infty }(\mathcal{F}_{1,0})$ with covariance function $\gamma
^{2}\Gamma $ and finite distribution characterized by (\ref{df}).
\end{theorem}

\begin{proof}
We begin by applying Corollary \ref{cor2}. For $T_{n}^{\ast
}(f)=(s(k/n)\sigma _{n}(f))^{-1}(T_{n}(f)-a_{n}(f)s(k/n))$ for $F\in
D(\Lambda )$\ or $T_{n}^{\ast }(f)=(a_{n}(f)/\sigma
_{n}(f))(T_{n}(f)/a_{n}-\gamma )$ for $F\in D(G_{1/\gamma }),$ we have 
\begin{equation*}
T_{n}^{\ast }(f)=V_{n}^{\ast }(f)+o_{P}^{\ast }(1),
\end{equation*}%
uniformly in $f\in \mathcal{F}_{h}$ and hence uniformly in $f\in \mathcal{F}%
_{h,0}$ $(h=1,2).$ From there, we apply Theorem 2.11.1 of (\cite{vaart}) on
the uniform behavior of the stochastic processes 
\begin{equation*}
V_{n}^{\ast }(f)=\sigma _{n}(f)^{-1}(V_{n}(f)-a_{n}(f))=\sum_{j=1}^{k}\frac{%
f(j)j^{-1}}{\sigma _{n}(f)}=:\sum_{j=1}^{k}Z_{j,n}(f),
\end{equation*}%
indexed by $f\in \mathcal{F}_{h,0}$. All the assumptions of Theorem 2.11.1
of (\cite{vaart}) have already been taken into account in our statement,
except this one, 
\begin{equation}
\mathbb{E}^{\ast }\{\left\Vert Z_{j,n}\right\Vert _{\mathcal{F}%
_{h}}^{2}I_{(\left\Vert Z_{j,n}\right\Vert _{\mathcal{F}_{h}}>\eta
)})\rightarrow 0,\text{ }as\text{ }n\rightarrow \infty ,  \tag{L3}
\end{equation}%
for any $\eta >0,$ and the convergence of the covariance function. But the $%
\left\Vert Z_{j,n}\right\Vert $ are measurable and 
\begin{equation*}
\mathbb{E}\left\Vert Z_{j,n}\right\Vert _{\mathcal{F}_{1}}=\max
\{f(j)/j,j\geq 1\}/\sigma _{n}(f)\rightarrow 0,
\end{equation*}%
because of $(KU1)$ and 
\begin{equation*}
\mathbb{E}\left\Vert Z_{j,n}\right\Vert _{\mathcal{F}_{2}}\leq
B_{n}\rightarrow 0
\end{equation*}%
because of $(KU2a)$. This proves $(L3)$. As to the covariance functions
which are 
\begin{equation*}
\Gamma _{n}(f_{1},f_{2})=\sum_{j=1}^{k(n)}\left\{ \sigma _{n}(f_{1})\sigma
_{n}(f_{1})\right\} ^{-1}f_{1}(j)f_{2}(j)j^{-2},
\end{equation*}%
we notice by the Cauchy-Schwartz inequality that they are bounded by the
unity. For $h=1,$ we have 
\begin{equation*}
\Gamma _{n}(f_{1},f_{2})=\frac{1}{\sqrt{A(2,f_{1})A(2,f_{2})}}%
\sum_{j=1}^{\infty }f_{1}(j)f_{2}(j)j^{-2}\leq 1,
\end{equation*}%
while for $h=2,$ the condition $(KU2a)$ guarantees the desired result. We
thus conclude that $\{T_{n}^{\ast }(f),f\in \mathcal{F}_{h,0}\}$ weakly
converges in $\ell ^{\infty }(\mathcal{F}_{h,0})$ for each $h=1,2.$ Now, by
Theorem \ref{theo2} and Corollary \ref{cor1}, we know that the weak limit is
either $\mathcal{L}$ defined by (\ref{df}) or a Gaussian process $\mathbb{G}$%
\ of covariance function $\Gamma $.
\end{proof}

\bigskip

\noindent Now, we present the second version which is more general since we
do not require the mesurability assumption so that we consider the whole
spaces $\mathcal{F}_{h}$ $(h=1,2).$

\begin{theorem}
\label{theo4} If $(\mathcal{F}_{h},\left\Vert {}\right\Vert )$ is a normed
space verifying the Riesz property, then the results of Theorem \ref{theo3}
hold when 
\begin{equation}
\int_{0}^{\delta _{n}}\sqrt{logN_{[]}(\epsilon ,\mathcal{F}_{h},L_{2}^{n})}%
d\epsilon \rightarrow 0,  \tag{L4}
\end{equation}%
as \ $n\rightarrow \infty $, in place of (L2), provided that the $Z_{j,n}$
have finite second moments
\end{theorem}

\begin{proof}
It is achieved by applying Theorem 2.11.9 of \cite{vaart} in the proof of
Theorem \ref{theo3}, that requires $(L1),$ $(L2),$ $(L4)$ and that the $%
Z_{j,n}$ have finite second moments. But $(L2)$ and the last condition hold.
Thus $(L1)$ and $(L4)$ together ensure the results of the theorem.
\end{proof}

\section{Special classes}

\label{sec5}

\noindent We specialize these results for the special class of the monotone
functions $f_{\tau }(j)=j^{\tau },$ $0<\tau .$ We will show here, in this
example, how to derive particular laws for special classes from our general
results. We know from \cite{demedioplo} and \cite{dioplo} that $%
T_{n}(f_{\tau })$ is asymptotically normal for $\tau >1/2$ while it
asymptotically follows a $\mathcal{L}(f_{\tau })$ type-law for $0<\tau <1/2$%
, under usual conditions of the $d.f.$ $G.$ We handle here the uniform
asymptotic behavior for these two range values of $\tau $ : $]0,1/2[$ and $%
[1/2,+\infty \lbrack .$ For the first case, we apply Theorem (\ref{theo3})
and for the second, Theorem \ref{theo4}. First, let $0<a<b<1/2$ and put$,$%
\begin{equation*}
\mathcal{F}_{0}(a,b)=\{f(j)=j^{\tau },a\leq \tau \leq b\}.
\end{equation*}

\noindent We have

\begin{corollary}
Let $0<a<b<1/2$ and $\mathcal{F}_{0}=\mathcal{F}_{0}(a,b)=\{f(j)=j^{\tau
},0<a\leq \tau \leq b<1/2\}.$ Then,

(1) if F$\in D(G_{0})$ and if (CU2) and (CU2) hold. Then 
\begin{equation*}
\left\{ \left(s(k/n)\sigma _{n}(f)\right)
^{-1}(T_{n}(f)-s(k/n)a_{n}(f)),f\in \mathcal{F}_{0}\right\}
\end{equation*}
weakly converges to $\{\mathcal{L}(f),f\in \mathcal{F}_{0}\}.$

(2) if $F\in D(G_{1/\gamma }),\gamma >0,$ and if (CU1) holds. Then 
\begin{equation*}
\left\{ (a_{n}/\sigma _{n}(f))(T_{n}(f)/a_{n}(f)-\gamma ),f\in \mathcal{F}%
_{0}\right\}
\end{equation*}%
weakly converges to $\{\gamma \mathcal{L}(f),f\in \mathcal{F}_{0}\}.$
\end{corollary}

\bigskip

\begin{proof}
We apply here Theorem \ref{theo3}. We have%
\begin{equation*}
\mathcal{F}_{0,1}=\{g(j)=j^{-(b-\tau )},a\leq \tau \leq b\}\text{.}
\end{equation*}%
For $f(j)=j^{\tau },$ denote $g_{f}(j)=j^{-(b-\tau )},$ that is 
\begin{equation*}
f(j)/j=g_{f}(j)\times j^{-(1-b)}.
\end{equation*}%
In this case, put 
\begin{equation*}
V_{n}^{\ast \ast
}(f)=(V_{n}(f)-a_{n}(f))=\sum_{j=1}^{k}g_{f}(j)(E_{j}-1)j^{-(1-b)}=:%
\sum_{j=1}^{k}Z_{j,n}(f).
\end{equation*}%
\noindent We have%
\begin{equation*}
\rho _{n}^{2}(f_{1},f_{2})=\mathbb{E}%
\sum_{j=1}^{k}(Z_{j,n}(f_{1})-Z_{j,n}(f_{2}))^{2}=%
\sum_{j=1}^{k}(f_{1}(j)-f_{2}(j))^{2}j^{-2}
\end{equation*}%
\begin{equation*}
\leq \rho ^{2}(f_{1},f_{2})=\sum_{j=1}^{\infty
}(f_{1}(j)-f_{2}(j))^{2}j^{-2}=\sum_{j=1}^{\infty
}(g_{f_{1}}(j)-g_{f_{2}}(j))^{2}j^{-2(1-b)}.
\end{equation*}

\noindent We point out that $\rho ^{2}(f_{1},f_{2})$ is nothing else but $%
\left\Vert g_{f_{1}}-g_{f_{2}}\right\Vert _{L_{2}(\mathcal{F}_{0,1},\mathbb{%
Q)}}^{2}$ for the probability measure on $\mathbb{N}$, 
\begin{equation*}
\mathbb{Q}=A(2,b)^{-1}\sum_{j=1}^{\infty }j^{-2(1-b)}\delta _{j}
\end{equation*}%
for%
\begin{equation*}
A(2,b)=\sum_{j=1}^{\infty }j^{-2(1-b)}<\infty ,
\end{equation*}%
\noindent For such monotone functions $g_{f}:\mathbb{N}\mapsto \lbrack 0,1]$%
, we have by virtue of Theorem 2.7.5 of \cite{vaart} that for some $K>0$,
any $\epsilon >0,$%
\begin{equation*}
N_{[]}(\epsilon ,\mathcal{F}_{0,1},L_{2}(\mathbb{Q}))\leq \exp (K\epsilon
^{-1}).
\end{equation*}%
This means that ($\mathcal{F}_{0},\rho )$ is totally bounded and $(L1)$ is
reduced to 
\begin{equation*}
\sup_{\rho (f_{1},f_{2})\leq \delta _{n}}\rho _{n}(f_{1},f_{2})\leq \delta
_{n}\rightarrow 0,
\end{equation*}%
\noindent which is trivial. In the same spirit%
\begin{equation*}
A(2,b,\omega )=\sum_{j=1}^{\infty }j^{-2(1-b)}(E_{j}-1)^{2}(\omega )
\end{equation*}%
\noindent is almost surely finite and%
\begin{equation*}
Q_{0}(\omega )=A(2,b,\omega )^{-1}\sum_{j=1}^{\infty
}j^{-2(1-b)}(E_{j}-1)^{2}(\omega )\delta _{j},
\end{equation*}%
\noindent is a probability measure for almost all $\omega $. And we have 
\begin{equation*}
0\leq d_{n}^{2}(f_{1},f_{2})\rightarrow d^{2}(f_{1},f_{2})=\left\Vert
g_{f_{1}}-g_{f_{2}}\right\Vert _{L_{2}(\mathcal{F}_{0,1},\mathbb{Q}_{0}%
\mathbb{)}}^{2}.
\end{equation*}%
This convergence is a continuity one since $\tau $ lies on the compact set $%
[a,b]$ (see Subsection \ref{ssec72}. in the Appendix \ref{sec7} for such
results)$.$ Thus, uniformly in $\tau \in $ $[a,b]$, for large values of $n,$%
\begin{equation*}
0\leq d_{n}^{2}(f_{1},f_{2})\geq 0.25\text{ }d^{2}(f_{1},f_{2}).
\end{equation*}%
\noindent We may use the same results of (\cite{vaart}) to get for some $K>0$
and for any $\epsilon >0,$%
\begin{equation*}
\log N_{[]}(\epsilon ,\mathcal{F}_{0},d_{n})\leq \log N_{[]}(\epsilon ,%
\mathcal{F}_{0},d/2)\leq K(\epsilon /2)^{-1}.
\end{equation*}%
This ensures $(L2).$ \noindent The covariance function is, for $%
f_{1}(j)=j^{\tau _{1}}$ and $f_{2}(j)=j^{\tau _{2}}$, 
\begin{equation*}
\Gamma ^{\ast }(f_{1},f_{2})=\lim_{n\rightarrow \infty
}\sum_{j=1}^{k}f_{1}(j)f_{2}(j)j^{-2}=\sum_{j=1}^{\infty }j^{-(2-\tau
_{1}-\tau _{2})}\leq \sum_{j=1}^{\infty }j^{-2(1-b)}=A(2,b).
\end{equation*}%
\noindent As to the mesurability hypothesis, it is readily seen that the
following supremum 
\begin{equation}
\sup_{(f_{1},f_{2})\in \mathcal{F}_{h},\text{ }\rho (f_{1},f_{2})\leq \delta
_{n}}\sum_{j=1}^{k}e_{j}\left\vert \sigma
_{n}^{-1}(f_{1})f_{1}(j)j^{-1}-\sigma
_{n}^{-1}(f_{2})f_{2}(j)j^{-1}\right\vert ^{p}E_{j}^{p}  \label{sup1}
\end{equation}%
is achieved through the rational values of $\tau $ in $[a,b]$, and then, is
measurable. This achieves the proof.
\end{proof}

\bigskip

\begin{corollary}
Let $1/2<a<b>1$ and $\mathcal{F}_{1}(a,b)=\{f(j)=j^{\tau },0<a\leq \tau \leq
b\}$.

(1) If F$\in D(G_{0})$ and if (CU3). Then 
\begin{equation*}
\left\{ \left( s(k/n)\sigma _{n}(f)\right)
^{-1}(T_{n}(f)-s(k/n)a_{n}(f)),f\in \mathcal{F}_{0}\right\}
\end{equation*}%
weakly converges to a Gaussian process $\mathbb{G}$ of covariance function $%
\Gamma $.

(2) If F$\in D(G_{1/\gamma }),\gamma >0,$ and if (CU1) holds. Then 
\begin{equation*}
\left\{ (a_{n}/\sigma _{n}(f))(T_{n}(f)/a_{n}(f)-\gamma ),f\in \mathcal{F}%
_{0}\right\}
\end{equation*}%
weakly converges to the Gaussian process $\gamma \mathbb{G}$.
\end{corollary}

\begin{proof}
We apply here Theorem \ref{theo4}. But, we begin by returning to the simple
scheme, that is, to fixed $f\in \mathcal{F}_{1}(a,b).$ Let $f(j)=j^{\tau }.$
We have 
\begin{equation*}
B(n,f)=O((\log k)^{-1}),
\end{equation*}%
$\tau =1/2,$%
\begin{equation*}
B(n,f)=O(k^{-(2\tau -1)})
\end{equation*}%
for $0<\tau <1/2$ and 
\begin{equation*}
B(n,f)=O(k^{-\tau }),
\end{equation*}%
for $\tau >1$. Then $(K1)$ holds and this leads to the normality case in
Theorem \ref{theo2} for each $\tau \geq 1/2.$ Next, for $f_{1}(j)=j^{\tau
_{1}}$ and $f_{2}(j)=j^{\tau _{2}},$ $\tau _{1}>1/2,$ $\tau _{2}>1/2,$%
\begin{equation*}
\lim_{n\rightarrow \infty }\sum_{j=1}^{k(n)}\left\{ \sigma _{n}(f_{1})\sigma
_{n}(f_{2})\right\} ^{-1}f_{1}(j)f_{2}(j)j^{-2}=\frac{\sqrt{(2\tau
_{1}-1)(2\tau _{2}-1)}}{\tau _{1}+\tau _{2}-1}=\Gamma (f_{1},f_{2})<\infty .
\end{equation*}%
And for $\tau _{1}=1/2$ and $\tau _{2}>1/2$%
\begin{equation*}
\sum_{j=1}^{k(n)}\left\{ \sigma _{n}(f_{1})\sigma _{n}(f_{1})\right\}
^{-1}f_{1}(j)f_{2}(j)j^{-2}\sim \sqrt{2\tau _{2}-1}(\log k)k^{(2\tau
_{2}-1)/2},
\end{equation*}%
that is, \noindent $\mathcal{F}_{1}(a,b)$ satisfies $(KU2a)$. This also
implies that we have the finite-distributions weak normality for $\tau >1/2.$
It also fullfils $(KU2b)$ since 
\begin{equation*}
B(n)=\sup_{f\in \mathcal{F}_{1}(a,b)}B(n,f)\leq k^{-b}\rightarrow 0.
\end{equation*}%
\noindent Consider here%
\begin{equation*}
V_{n}^{\ast \ast }(f)=\sigma
_{n}(f)^{-1}(V_{n}(f)-a_{n}(f))=\sum_{j=1}^{k}f(j)/(j\sigma
_{n}(f))(E_{j}-1)=:\sum_{j=1}^{k}Z_{j,n}(f)
\end{equation*}%
We have%
\begin{equation*}
\rho _{n}^{2}(f_{1},f_{2})=\mathbb{E}%
\sum_{j=1}^{k}(Z_{j,n}(f_{1})-Z_{j,n}(f_{2}))^{2}=2-\frac{2}{\sigma _{n}(f{1}%
)\sigma _{n}(f_{2})}\sum_{j=1}^{k}f_{1}(j)f_{2}(j)j^{-2}
\end{equation*}%
\begin{equation*}
\rightarrow 2(1-\Gamma (f_{1},f_{2}))=\sum_{j=1}^{\infty }\left\{
(f_{1}(j)/(j\sigma _{n}(f_{1}))-(f_{2}(j)/(j\sigma _{n}(f_{2}))\right\} ^{2}.
\end{equation*}

\noindent By routine calculations and by a continuity argument based on the
remark that our $\tau ^{\prime }s$ are in the compact set $[a,b]$, we may
show that%
\begin{equation}
\rho _{n}^{2}(f_{1},f_{2})<3(1-\Gamma (f_{1},f_{2}))  \label{cont}
\end{equation}

\noindent for large values of $n$. Since we use this type of arguments many
times, we show in Subsection \ref{ssec72} of the appendix (\ref{sec7}) the
exact proof of (\ref{cont}). But%
\begin{equation*}
1-\Gamma (f_{1},f_{2})=(1-\frac{\sqrt{(2\tau _{1}-1)(2\tau _{2}-1)}}{\tau
_{1}+\tau _{2}-1})
\end{equation*}%
\noindent For $\tau _{1}-\tau _{2}=\delta >0,$%
\begin{equation*}
1-\Gamma (f_{1},f_{2})=\frac{(2\tau _{2}-1)+\delta -\sqrt{(2\tau
_{2}-1)^{2}+2\delta (2\tau _{2}-1)}}{\tau _{1}+\tau _{2}-1}.
\end{equation*}

\noindent We use a Taylor expansion of a first order, to get for $1/2<a<\tau
_{1},\tau _{2}<b,$ $3(1-\Gamma (f_{1},f_{2}))\leq B(a,b)\delta ,$ \noindent
where $B(a,b)$ depend only on $a$ and $b$. Thus for a fixed $\delta $, for
large values of $k$,%
\begin{equation*}
\rho _{n}^{2}(f_{1},f_{2})\leq B(a,b)\delta.
\end{equation*}

\noindent We may now take the metric $\rho (f_{1},f_{2})=\Vert \tau
_{1}-\tau _{2}\Vert $, for which $(\mathcal{F}_{1}(a,b),\rho _{2})$ is a
Riesz Space totally bounded and we surely obtain 
\begin{equation*}
\sup_{\rho _{2}(f_{1},f_{2})\leq \delta _{n}}\rho _{n}(f_{1},f_{2})\leq
4B(a,b)\delta _{n}\rightarrow 0.
\end{equation*}%
This gives the $(L1)$ hypothesis. Now, to conclude the proof by establishing
the functional law as already described in Corollary \ref{cor1}, we have to
prove that $(L4)$ holds for $(\mathcal{F}_{1}(a,b),\rho _{2})$ with
partitions not depending on $n$, so that $(L2)$ is unnecessary. Since the
proof concerning $(L4)$ is very technical, we state it Subsection (\ref%
{ssec71}) of the Appendix (\ref{sec7}).
\end{proof}

\section{Technical lemmas}

\label{sec6}

\label{sectiontool}Define the following conditions

\begin{equation}
A(2,f)=\sum_{j=1}^{\infty }f(j)^{2}j^{-2}\in ]0,+\infty \lbrack ,  \tag{K1}
\end{equation}
with the notation (well-defined since $f>0$) 
\begin{equation*}
A(n,f)=\sum_{j=1}^{\infty }f(j)^{n}j^{-n}
\end{equation*}
\noindent and 
\begin{equation}
\limsup_{n \rightarrow +\infty} \sigma _{n}(f)^{-1}\max \{f(j)j^{-1},1\leq
j\leq k\}= \limsup_{n \rightarrow +\infty} B(n,f) = 0  \tag{K2}
\end{equation}

\noindent We begin by this simple lemma where we suppose that we are given a
sequence of independent and uniformly distributed random variables $%
U_{1},U_{2},...$ as in (\ref{repre}).

\begin{lemma}
\label{lemmatool} \label{lemma1} Let 
\begin{equation*}
V_{n}(f)=\sum_{j=1}^{k}f(j)\log (\frac{U_{j+1,n}}{U_{j,n}}).
\end{equation*}
If $(K2)$ \ holds, 
\begin{equation*}
\sigma _{n}^{-1}(f)(V_{n}(f)-a_{n}(f))\leadsto \mathcal{N}(0,1)
\end{equation*}
and if $(K1)$ holds$,$%
\begin{equation*}
\sigma _{n}^{-1}(f)(V_{n}(f)-a_{n}(f))\leadsto \mathcal{L}(f),
\end{equation*}
where $a_{n}(f)$ and $\sigma _{n}(f)$ are defined in (\ref{defan}) and 
\begin{equation*}
\mathcal{L}(\tau )=A(2,f)\sum_{j=1}^{\infty }f(j)j^{-1}(E_{j}-1),
\end{equation*}
is a centred and reduced random variable with all finite moments.
\end{lemma}

\begin{proof}
By using the Malmquist representation (\ref{malm}), we have 
\begin{equation*}
V_{n}(f)=\sum_{j=1}^{k} f(j)j^{-1}E_{j}.
\end{equation*}
It follows that $\mathbb{E}(V_{n}(f))=a_{n}(f)$ and $\mathbb{V}%
ar(V_{n}(f))=\sigma _{n}^{2}(f).$ Put

\begin{equation*}
V_{n}^{\ast }(f)=\sigma _{n}(f)^{-1}(V_{n}(f)-a_{n}(f)).
\end{equation*}
Then

\begin{equation*}
V_{n}^{\ast }(f)=\sigma _{n}(f)^{-1}\sum_{j=1}^{k}f(j)j^{-1}(E_{j}-1).
\end{equation*}
First suppose that (K1) holds, that is $\sigma _{n}(f)\rightarrow
A(2,f)^{-1/2}\in ]0,1[.$ Then 
\begin{equation*}
V_{n}^{\ast }(f)\rightarrow A(2,f)^{-1/2}\sum_{j=1}^{\infty
}f(j)j^{-1}(E_{j}-1)=\mathcal{L}(f).
\end{equation*}
Now, we have to prove that $\mathcal{L}(f)$ is a well-defined random
variable with all finite moments. The moment characteristic function of $%
V_{n}^{\ast }(f)$ is 
\begin{equation*}
\psi _{V_{n}^{\ast }(f)}(t)=\exp
(-A(2,f)^{-1/2}\sum_{j=1}^{k}f(j)j^{-1}(it))\prod_{j=1}^{k}(1-it\times
f(j)j^{-1}A(f)^{-1/2})^{-1}.
\end{equation*}
By using the development of $\log (1-\cdot ),$ and, by the Lebesgues
Theorem, one readily proves that 
\begin{equation*}
\psi _{V_{n}^{\ast }(\tau )}(t)=\exp (\sum_{j=1}^{k}\sum_{n=2}^{\infty }%
\frac{(it)^{n}}{n}f(j)^{n}j^{-n}A(2,f)^{-n/2})
\end{equation*}
\begin{equation}
\rightarrow \psi _{\infty }(t)=\exp (\sum_{n=2}^{\infty }\frac{(it)^{n}}{n}%
A(n,f)A(2,f)^{-n/2}).  \label{fc01}
\end{equation}
We note that if $A(2,f)<\infty ,$ then $A(n,f)$ is also finite for any $%
n\geq 2$, since the general term (in $j$) of $A(n,f)$ is less than that of $%
A(2,f)$, for large values of $j$. This concludes the proof when $(K1)$ holds.%
\newline

\noindent Now suppose that $(K2)$ holds. Let us evaluate the moment
generating function of $V_{n}^{\ast }(f):$

\begin{equation}
\phi_{V_{n}^{\ast }(f)}(t)=\prod_{j=1}^{k}\phi
_{(E_{j}-1)}(tf(j)j^{-1}\sigma _{n}(f)^{-1}).  \label{tfl2}
\end{equation}

\noindent Recall that, in this case, $\sigma _{n}(f)\uparrow \infty$. It
follows from $(K2)$, that for any $u_{0}>0$\ for a fixed $t$, for $k$ large
enough, 
\begin{equation}
\left| t\text{ }f(j)j^{-1}\text{ }\sigma _{n}(f)^{-1}\right| \leq u_{0}
\label{dl10c}
\end{equation}
uniformly in $j\geq 1$. At this step, we use the expansion of $\phi
_{(E_{j}-1)}$ in the neighborhood of zero : 
\begin{equation*}
\psi _{(E_{j}-1)}(u)=1+u^{2}/2+u^{3}g(u),
\end{equation*}
where there exists $u_{0}$ such that 
\begin{equation*}
0\leq u\leq u_{0}\Rightarrow \left| g(u)\right| \leq 1.
\end{equation*}
Using the uniform bound in (\ref{dl10c}), we get 
\begin{equation}
\phi _{(E_{j}-1)}(t\text{ }f(j)j^{-1}\text{ }\sigma _{n}(f)^{-1})=1+\frac{1}{%
2}(t\text{ }f(j)j^{-1}\text{ }\sigma _{n}(f))^{-1})^{2}  \label{tfl1}
\end{equation}
\begin{equation*}
+(t\text{ }f(j)j^{-1}\text{ }\sigma _{n}(f))^{-1})^{3}g_{0,j,n}(t),
\end{equation*}
where $\left| g_{0,j,n}(t)\right| \leq 1$ for all $1\leq j\leq k.$ By the
uniform boundedness of the error term, we have 
\begin{equation*}
\log \phi_{(E_{j}-1)}(t\text{ }f(j)j^{-1}\sigma _{n}(f))^{-1})=
\end{equation*}
\begin{equation*}
\frac{1}{2}(t\text{ }f(j)j^{-1}\text{ }\sigma _{n}(f))^{-1})^{2}+(t\text{ }%
f(j)j^{-1}\sigma _{n}(f))^{-1})^{3}g_{0,j,n}(t)
\end{equation*}
\begin{equation*}
+(t\text{ }f(j)j^{-1}\text{ }\sigma _{n}(f))^{-1})^{3}g_{1,j,n}(t),
\end{equation*}
where, always $\left| g_{1,j,n}(t)\right| \leq 1$ for all $1\leq j\leq k.$
Finally 
\begin{equation*}
\phi _{V_{n}^{\ast }(f)}(t)=\exp (\sum_{j=1}^{k}\log \phi _{(E_{j}-1)}(t%
\text{ }f(j)j^{-1}\sigma _{n}(f))^{-1})
\end{equation*}
\begin{equation*}
=\exp (t^{2}/2+g_{2,j,n}(t)\times t^{3}\sigma
_{n}(f)^{-3}\sum_{j=1}^{k}f(j)^{3}j^{-3}),
\end{equation*}
with $\left| g_{1,j,n}(t)\right| \leq 2$ for all $1\leq j\leq k$. Since 
\begin{equation*}
0\leq \sigma _{n}(f)^{-3}\sum_{j=1}^{k}f(j)^{3}j^{-3})\leq B(n,f)\times
\sigma _{n}(f)^{-2}\sum_{j=1}^{k}f(j)^{2}j^{-2}=B(n,f)\rightarrow 0.
\end{equation*}
Hence 
\begin{equation*}
\phi _{V_{n}^{\ast }(f)}(t)\rightarrow \exp (t^{2}/2)
\end{equation*}
and 
\begin{equation*}
V_{n}^{\ast }(f)\rightarrow \mathcal{N}(0,1).
\end{equation*}
\end{proof}

\begin{lemma}
\label{lemma2} Let $(a,b)\in \mathbb{R}^{2}$ and suppose that for any couple 
$(f_{1},f_{2})\in \mathcal{F}_{h}^{2}$, $h=1,2,$, 
\begin{equation*}
\lim_{n\rightarrow \infty }\frac{1}{\sigma _{n}(f_{1})\sigma _{n}(f_{2})}%
\sum_{j=1}^{k}f_{1}(j)f_{2}(j)j^{-2}=\Gamma(f_{1},f_{2})\text{ exists.}
\end{equation*}
Then for $h=1$, $aV_{n}^{\ast }(f_{1})+bV_{n}^{\ast }(f_{2})$ weakly
converges to 
\begin{equation*}
a\mathcal{L}(f_{1})+b\mathcal{L}(f_{2}).
\end{equation*}
and for h=2, $aV_{n}^{\ast }(f_{1})+bV_{n}^{\ast }(f_{2})$ weakly converges
to a normal random of variance : 
\begin{equation*}
v(a,b,f_{1},f_{2})=a^{2}+b^{2}+2ab\Gamma _{1}(f_{1},f_{2}).
\end{equation*}
In both cases, the finite-distributions of \{$V_{n}^{\ast }(f),f\in \mathcal{%
F}_{h}\}$ weakly converge to those of the process $\mathcal{L}$ for $h=1$
and, for $h=2,$ to those of a Gaussian process of covariance function 
\begin{equation*}
\Gamma (f_{1},f_{2})=\lim_{n\rightarrow \infty }\frac{1}{\sigma
_{n}(f_{1})\sigma _{n}(f_{2})}\sum_{j=1}^{k}f_{1}(j)f_{2}(j)j^{-2}
\end{equation*}
provived these numbers are finite.
\end{lemma}

\begin{proof}
The case $h=1$ is straightforward. For $h=2$, we slightly change the proof
of the previous lemma from (\ref{tfl2}). Put $%
V_{n}(a,b,f_{1},f_{2})=aV_{n}^{\ast }(f_{1})+bV_{n}^{\ast }(f_{2})$ and 
\begin{equation*}
c(n,j)=\sigma _{n}(f_{1})^{-1}af_{1}(j)j^{-1}+\sigma
_{n}(f_{2})^{-1}bf_{2}(j)j^{-1}.
\end{equation*}
We alway have, for $(f_{1},f_{2})\in \mathcal{F}_{2}^{2},$%
\begin{equation*}
\sup_{j}\left| c(n,j)\right| \rightarrow 0
\end{equation*}
and 
\begin{equation*}
\sum_{j=1}^{k}c(n,j)^{2}=a^{2}+b^{2}+2ab\frac{1}{\sigma _{n}(f_{1})\sigma
_{n}(f_{2})}\sum_{j=1}^{k}f_{1}(j)f_{2}(j)j^{-2}
\end{equation*}
\begin{equation*}
\rightarrow v(a,b,f_{1},f_{2})=a^{2}+b^{2}+2ab\Gamma(f_{1},f_{2}).
\end{equation*}
By using the same arguments in (\ref{tfl2}) and (\ref{tfl1}), we have 
\begin{equation*}
\phi _{V_{n}(a,b,f_{1},f_{2})}(t)=\exp (\sum_{j=1}^{k}\log \psi
_{(E_{j}-1)}(t\text{ }c(n,j))
\end{equation*}
\begin{equation*}
=\exp (\left\{ \sum_{j=1}^{k}c(n,j)^{2}\right\} t^{2}/2+g_{3,j,n}(t)\times
t^{3}(\sum_{j=1}^{k}c(n,j)^{3})),
\end{equation*}
with $\left| g_{3,j,n}(t)\right| \leq 2$ for all $1\leq j\leq k.$ Since 
\begin{equation*}
0\leq \sum_{j=1}^{k}c(n,j)^{3}\leq \left\{ aB(n,f_{1})+bB(n,f_{2})\right\}
\times \sum_{j=1}^{k}c(n,j)^{2}\rightarrow 0,
\end{equation*}
we get 
\begin{equation*}
\phi _{V_{n}(a,b,f_{1},f_{2})}(t)\rightarrow \exp
(v(a,b,f_{1},f_{2})t^{2}/2).
\end{equation*}
This achieves the proof. Now, these methods are reproducible for any finite
linear combination 
\begin{equation*}
a_{1}V_{n}^{\ast }(f_{1})+a_{2}V_{n}^{\ast }(f_{2})+...+a_{m}V_{n}^{\ast
}(f_{m}).
\end{equation*}
In all cases, we find the same finite distribution laws. For $h=1,$ $%
(V_{n}^{\ast }(f_{1}),...,V_{n}(f_{m}))$ converges in law to $(\mathcal{L}%
(f_{1}),...,\mathcal{L}(f_{m}))$ and for $h=2,$ $(V_{n}^{\ast
}(f_{1}),...,V_{n}^{\ast }(f_{m}))$ converges to a Gaussian vector of
variance-covariance matrix $(\Gamma (f_{i},f_{j}),1\leq i,j\leq m)$ provided
these numbers are finite.
\end{proof}

\section{Appendix}

\label{sec7}

\subsection{Check hypothesis (L4) for $\protect\tau >1/2$}

\label{ssec71}

\bigskip

We shall consider $\mathcal{F}_{2}(a,b)$ as a normed vector by identifying $%
f_{\tau }(j)=j^{\tau }$ with $\tau $ and setting $f_{\tau _{1}}+f_{\tau
_{2}}=f_{\tau _{1}+\tau _{2}}$ and $f_{\tau _{1}}\leq $ $f_{\tau _{2}}$ $iff$
$\ \tau _{1}\leq \tau _{2}.$ We put without lost of generality that $b-a=1$
and that $\varepsilon ^{2}=1/p$. Next we devide $\mathcal{F}_{2}(a,b)$ into $%
p$ intervals $[0,\tau _{1}[,[\tau _{1},\tau _{2}[,...,[\tau _{p-1},\tau
_{p}].$ Now, put%
\begin{equation*}
\sigma _{n}^{2}(1,\tau )=k^{-2\tau +1}\sigma _{n}^{2}(f_{\tau })
\end{equation*}%
and consider $\tau _{i}<\nu <\tau <\tau _{i+1}$ and let $i=1$ for short, $%
\tau -\nu =\delta $ and

\begin{equation}
d_{n}^{2}(f_{\nu },f_{\tau },j)=(f_{\nu }(j)/\sigma _{n}(f_{\nu })-f_{\tau
}(j)/\sigma _{n}(f_{\tau }))^{2}j^{-2}(E_{j}-1)^{2}  \label{L00a}
\end{equation}%
\begin{equation*}
\left\{ \frac{k^{-2\nu +1}}{\sigma _{n}^{2}(1,\nu )}j^{2\nu -2}+\frac{%
k^{-2\tau +1}}{\sigma _{n}^{2}(1,\nu )}j^{2\tau -2}-2\frac{k^{-\nu -\tau +1}%
}{\sigma _{n}(1,\nu )\sigma _{n}(1,\tau )}j^{\tau +\nu -2}\right\}
(E_{j}-1)^{2}
\end{equation*}%
\begin{equation*}
=\left\{ T_{0}(n,\tau ,j)+T_{0}(n,\nu ,j)-2T_{1}(n,\nu ,\tau )\right\}
(E_{j}-1)^{2}.
\end{equation*}%
Let us handle $T_{0}(n,\tau ,j).$ By adding to it this null term 
\begin{equation*}
0=\left\{ -\sigma _{n}^{-2}(1,\nu )k^{-2\tau +1}j^{2\tau -2}-\sigma
_{n}^{-2}(1,\nu )k^{-2\tau +1}j^{2\tau -2}\right\}
\end{equation*}
\begin{equation*}
+\{-\sigma _{n}^{-2}(1,\nu )k^{-2\nu +1}j^{2\nu -2}+-\sigma _{n}^{-2}(1,\nu
)k^{-2\nu +1}j^{2\nu -2}\},
\end{equation*}
we get 
\begin{equation}
T_{0}(n,\tau ,j)=(\frac{1}{\sigma _{n}^{2}(1,\tau )}-\frac{1}{\sigma
_{n}^{2}(1,\nu )})k^{-2\tau +1}j^{2\tau -2}  \label{L01a}
\end{equation}%
\begin{equation}
+\frac{k^{-2\nu +1}}{\sigma _{n}^{2}(1,\nu )}j^{2\nu -2}(1-(j/k)^{2\delta })+%
\frac{k^{-2\nu +1}}{\sigma _{n}^{2}(1,\nu )}j^{2\nu -2},  \label{L01b}
\end{equation}
where we used in the previous line (\ref{L01a}) the following identity%
\begin{equation}
k^{-2\tau +1}j^{2\tau -2}=k^{-2\nu +1}j^{2\nu -2}+k^{-2\nu +1}j^{2\nu
-2}(1-(j/k)^{2\delta }).  \label{L02}
\end{equation}%
Using again (\ref{L02}) in (\ref{L01a}), we arrive at%
\begin{equation}
T_{0}(n,\tau ,j)
\end{equation}

\begin{equation*}
=T_{0}(n,\nu ,j)+(\frac{1}{\sigma _{n}^{2}(1,\tau )}-\frac{1}{\sigma
_{n}^{2}(1,\nu )})k^{-2\nu +1}j^{2\nu -2} \label{L03c} 
\end{equation*}
\begin{equation*}
+(\frac{1}{\sigma _{n}^{2}(1,\tau )}-\frac{1}{\sigma _{n}^{2}(1,\nu )}%
)k^{-2\nu +1}j^{2\nu -2}(1-(j/k)^{2\delta })+\frac{k^{-2\nu +1}}{\sigma
_{n}^{2}(1,\nu )}j^{2\nu -2}(1-(j/k)^{2\delta }).
\end{equation*}%
We use the same techniques to also get%
\begin{equation}
T_{1}(n,\nu ,\tau )
\end{equation}

\begin{equation*}
=T_{0}(n,\nu ,j)+(\frac{1}{\sigma _{n}(1,\tau )\sigma _{n}(1,\nu )}-\frac{1}{%
\sigma _{n}^{2}(1,\nu )})k^{-2\nu +1}j^{2\nu -2} 
\end{equation*}
\begin{equation*}
+(\frac{1}{\sigma _{n}(1,\tau )\sigma _{n}(1,\nu )}-\frac{1}{\sigma
_{n}^{2}(1,\nu )})k^{-2\nu +1}j^{2\nu -2}(1-(j/k)^{\delta })+\frac{k^{-2\nu
+1}}{\sigma _{n}^{2}(1,\nu )}j^{2\nu -2}(1-(j/k)^{\delta })  \label{L03b}
\end{equation*}%
Let us handle the $\sigma _{n}^{-2}(1,\tau )-\sigma _{n}^{-2}(1,\nu)$ in (%
\ref{L03c}). We get, by using the same methods, 
\begin{equation*}
\sigma _{n}^{-2}(1,\tau )-\sigma _{n}^{-2}(1,\nu )=\sigma _{n}^{-2}(1,\tau
)\sigma _{n}^{-2}(1,\tau )k^{-2\nu +1}\sum_{j=1}^{k}j^{2\nu
-2}(1-(j/k)^{2\delta }).
\end{equation*}%
We already noticed that $\sigma _{n}^{2}(1,\tau )\rightarrow (2\tau
-1)^{-1}\in \lbrack B=(2b-1)^{-1},A=(2a-1)^{-1}]$ uniformly in $\tau \in
\lbrack a,b]$ by a continuity convergence argument. For $\varepsilon $
chosen such that $2a-1-2\varepsilon >0,$ we get%
\begin{equation*}
k^{-2\nu +1}\sum_{j=1}^{k}j^{2\nu -2}(1-(j/k)^{2\delta })\leq
\sum_{j=1}^{k}j^{-1}(j/k)^{2a-1}(1-(j/k)^{2\varepsilon ^{2}})
\end{equation*}%
\begin{equation*}
=k^{2a-1}\sum_{j=1}^{k}j^{2a-2}-k^{2\varepsilon ^{2}}j^{2a-2+2\varepsilon
^{2}}.
\end{equation*}%
By using the asymptotic approximations to the corresponding integrals, we
get, as $n\rightarrow \infty ,$%
\begin{equation*}
k^{2a-1}\sum_{j=1}^{k}j^{2a-2}-k^{2\varepsilon ^{2}}j^{2a-2+2\varepsilon
^{2}}
\end{equation*}%
\begin{equation*}
=k^{-2a+1}K\{(1+o(1))\frac{k^{2a-1}-1}{2a-1}\}-(1+o(1)k^{-2\varepsilon ^{2}}%
\frac{k^{2a-1+2\varepsilon ^{2}}}{2a-1+2\varepsilon ^{2}})
\end{equation*}%
\begin{equation*}
\rightarrow \{(2a-1)^{-1}-(2a-1+2\varepsilon ^{2})^{-1}\}\leq (2\varepsilon
^{2})/(2a-1)^{2}.
\end{equation*}%
We conclude that, for large value of $n,$%
\begin{equation}
k^{-2\nu +1}\sum_{j=1}^{k}j^{2\nu -2}(1-(j/k)^{2\delta })\leq K\varepsilon
^{2}  \label{L04}
\end{equation}%
and hence%
\begin{equation*}
\left\vert \sigma _{n}^{-2}(1,\tau )-\sigma _{n}^{-2}\right\vert \leq
K\varepsilon ^{2},
\end{equation*}%
for some positive universal constant $K.$ This constant is generic and may
change from line to line. In the same sipirit, we have, for this term in (%
\ref{L03b}),%
\begin{equation*}
\left\{ \sigma _{n}(1,\tau )\sigma _{n}(1,\nu )\right\} ^{-1}-\sigma
_{n}^{-2}(1,\nu )=\sigma _{n}(1,\nu )^{-1}\sigma _{n}(1,\tau )^{-1}-\sigma
_{n}^{-1}(1,\nu )
\end{equation*}%
\begin{equation*}
=-\frac{\sigma _{n}^{-2}(1,\tau )-\sigma _{n}^{-2}(1,\nu )}{\sigma
_{n}(1,\nu )^{2}\sigma _{n}(1,\tau )(\sigma _{n}(1,\tau )+\sigma _{n}(1,\nu
))}.
\end{equation*}%
The methods used above lead to%
\begin{equation}
\left\vert \left\{ \sigma _{n}(1,\tau )\sigma _{n}(1,\nu )\right\}
^{-1}-\sigma _{n}^{-2}(1,\nu )\right\vert \leq K\varepsilon ^{2}.
\label{L05}
\end{equation}%
Now returning to Formalae (\ref{L00a}), (\ref{L01a}) and (\ref{L03c}), we
see that the terms $T_{0}(n,\nu ,j)$ disappears as $T_{0}(n,\nu
,j)-2T_{0}(n,\nu ,j)+T_{0}(n,\nu ,j)=0.$ Computing the expectation of the
remainder terms, we get

\bigskip 
\begin{equation*}
Ed_{n}^{2}(f_{\nu },f_{\tau },j)\leq 2\left\vert \frac{1}{\sigma
_{n}^{2}(1,\tau )}-\frac{1}{\sigma _{n}^{2}(1,\nu )}\right\vert
j^{-1}(j/k)^{2a-1}
\end{equation*}

\bigskip 
\begin{equation*}
+2\left\vert \frac{1}{\sigma _{n}(1,\tau )\sigma _{n}(1,\nu )}-\frac{1}{%
\sigma _{n}^{2}(1,\nu )}\right\vert
j^{-1}(j/k)^{2a-1}+2j^{-1}(j/k)^{2a-1}(1-(j/k)^{2\delta }).
\end{equation*}%
The left-member does not depend on $(\nu ,\tau )\in \lbrack \tau _{i-1},\tau
_{i}]$ so that, with the application (\ref{L04}) and (\ref{L05}), we obtain 
\begin{equation*}
\max_{(\nu ,\tau )\in \lbrack \tau _{i-1},\tau _{i}]} d_{n}^{2}(f_{\nu
},f_{\tau },j)\leq K\left\{ (j/k)^{2a-1}\varepsilon
^{2}+2j^{-1}(j/k)^{2a-1}(1-(j/k)^{2\varepsilon ^{2}})\right\}(E_{j}-1)^{2} .
\end{equation*}%
Finally, by (\ref{L04}),%
\begin{equation*}
\sum_{j=1}^{k} \mathbb{E} \sup_{(\nu ,\tau )\in \lbrack \tau _{i-1},\tau
_{i}]} d_{n}^{2}(f_{\nu },f_{\tau },j)\leq K\varepsilon ^{2},
\end{equation*}%
for large values of $n$. Then 
\begin{equation*}
logN_{[]}(\epsilon ,\mathcal{F}_{h},L_{2}^{n}) \leq p \leq K\varepsilon
^{-2}. 
\end{equation*}
\noindent And this ensures $(L4)$ of Theorem \ref{theo4}. \bigskip

\subsection{Continuous convergence.}

\label{ssec72}

\noindent In order to prove (\ref{cont}), suppose for each $N>0$, there
exists a value $n \geq N$ and a couple $(f_{1},f_{2})$ such that

\begin{equation}
\rho _{n}^{2}(f_{1},f_{2})<(1-\Gamma (f_{1},f_{2}))  \tag{H}
\end{equation}

\noindent We may lessen the notations and put $\rho_{n}
^{2}(f_{1},f_{2})=\rho_{n} ^{2}(\tau_{1},\tau_{2})$ for $f_{1}(j)=j^{%
\tau_{1}}$ and $f_{2}(j)=j^{\tau_{2}} \in (a,b)^{2}$. It is easy to prove
that

\begin{equation*}
\rho _{n}^{2}(\tau _{1},\tau _{2})\rightarrow 2(1-\Gamma (\tau _{1},\tau
_{2})),
\end{equation*}

\noindent continuously, that is, 
\begin{equation*}
\rho _{n}^{2}(\tau _{1,n},\tau _{2,n})\rightarrow 2(1-\Gamma (\tau _{1},\tau
_{2})).
\end{equation*}

if $(\tau_{1,n},\tau_{2,n}) \rightarrow(\tau_{1},\tau_{2})$, as $%
n\rightarrow +\infty$. But, with our hypothesis (H), we can find a sequence
of integers $n_{1} < n_{2} < ... < n_{k} < n_{k+1} < .. $ such that and a
sequence of couples $(\tau_{1,n_{k}},\tau_{2,n_{k}}) \in (a,b)^{2}$, $k=1,2,
...$ such that for any $k$,

\begin{equation}
\rho _{n}^{2}(\tau _{1,n_{k}},\tau _{2,n_{k}})<(1-\Gamma (\tau
_{1,n_{k}},\tau _{2,n_{k}}).  \tag{H1}
\end{equation}

\noindent By the Bolzano-Weierstrass Theorem, we may extract from $(\tau
_{1,n_{k}},\tau _{2,n_{k}})$ a subsequence, denoted $(\tau _{1,n_{k}^{\ast
}},\tau _{2,n_{k}^{\ast }})$ converging to some $(\tau _{1},\tau _{2})\in
(a,b)^{2}$ and by the continuity result 
\begin{equation*}
\rho _{n^{\ast }}^{2}(\tau _{1,n_{k}^{\ast }},\tau _{2,n_{k}^{\ast
}})\rightarrow 2(1-\Gamma (\tau _{1},\tau _{2})).
\end{equation*}

\noindent This violates $(H1)$ and then proves (\ref{cont}).

\end{document}